\documentclass[12pt]{article}

\usepackage{graphicx}
\usepackage{latexsym}
\usepackage{amssymb}
\usepackage{array}

\def\0{\over } \def\1{\vec } \def\2{{1\over2}} \def\4{{1\over4}}
\def\5{\bar } 
\def\6{\partial }
\def\7#1{{#1}\llap{/}}
\def\8#1{{\textstyle{#1}}} \def\9#1{{\bf{#1}}}
\def\.{\dot }
\def\^#1{\widehat{#1}}
\def\ket#1{|#1\rangle }

\def\({\left(} \def\){\right)} \def\<{\left\langle } \def\>{\right\rangle }
\def\[{\left[} \def\]{\right]}  
 
\def\pmbf#1{\setbox0=\hbox{${#1}$}
        \kern-.025em\copy0\kern-\wd0
        \kern.05em\copy0\kern-\wd0
        \kern-.025em\raise.0433em\box0 }

\def\be{\begin{equation}}
\def\ee{\end{equation}}

\newcommand{\bel}[1]{\begin{equation}\label{#1}}
\def\bea{\begin{eqnarray}}
\newcommand{\beal}[1]{\begin{eqnarray}\label{#1}}
\def\eea{\end{eqnarray}}
\def\nn{\nonumber\\ }

\newcommand{\pv}{\phi_{\mathrm V}}

\newcommand{\SUSY}{susy}

\begin{document}

\begin{titlepage}
\renewcommand{\thefootnote}{\alph{footnote}}
~\vspace{-3cm}
\begin{flushright}
TUW-03-17\\
YITP-SB-03-37\\
ESI 1344
\end{flushright}  
\vfil
\centerline{\large\bf Nonvanishing quantum corrections to the mass and }
\smallskip
\centerline{\large\bf central charge of the $N=2$ vortex and BPS saturation}
\begin{center}
\vfil 
{\large  
A. Rebhan$^1$\footnote{\footnotesize\tt rebhana@hep.itp.tuwien.ac.at}, 
P. van Nieuwenhuizen$^{2,3}$\footnote{\footnotesize\tt vannieu@insti.physics.sunysb.edu} 
and R. Wimmer$^1$\footnote{\footnotesize\tt rwimmer@hep.itp.tuwien.ac.at}
}\\
\end{center}  \medskip \smallskip \qquad \qquad 
{\sl $^1$} \parbox[t]{12cm}{ \sl 
  Institut f\"ur Theoretische Physik, Technische Universit\"at Wien, \\
  Wiedner Hauptstr. 8--10, A-1040 Vienna, Austria\\ } \\
\bigskip \qquad \qquad 
{\sl $^2$} \parbox[t]{12cm}{ \sl 
  C.N.Yang Institute for Theoretical Physics, \\
  SUNY at Stony Brook, Stony Brook, NY 11794-3840, USA } \\
\bigskip \qquad \qquad 
{\sl $^3$} \parbox[t]{12cm}{ \sl 
  Erwin Schr\"odinger Int.\ Institute for Mathematical Physics, \\
   Boltzmanngasse 9, A-1090 Vienna, Austria\\ } \\
\vfil
\centerline{ABSTRACT}\vspace{.5cm}
  \begin{quotation}
The one-loop quantum corrections to the mass and central charge
of the $N=2$ vortex in 2+1 dimensions are determined
using supersymmetry-preserving
dimensional regularization by dimensional reduction of the corresponding
$N=1$ model with Fayet-Iliopoulos term in 3+1 dimensions.
Both the mass and the central charge turn out to have
nonvanishing one-loop corrections which however are equal
and thus saturate the Bogomolnyi bound.
We explain BPS saturation by standard multiplet shortening arguments,
correcting a previous claim in the literature
postulating the presence of a second degenerate short multiplet
at the quantum level.
  \end{quotation}
\vfil
\end{titlepage}

\setcounter{footnote}{0}

\section{Introduction and summary}

In supersymmetric (\SUSY) theories with solitons, topological quantum
numbers appear as central extensions of the \SUSY\ algebra.
In the topological sectors there is then typically a lower bound to the mass
spectrum determined by the central charge, and BPS 
(Bogomolnyi-Prasad-Sommerfield) states, which
saturate this lower bound, form shortened multiplets when
compared to the usual massive multiplets. They
are of particular interest because this ``multiplet shortening''
ties the mass of these states to their topological quantum numbers
such that the BPS saturation is protected from quantum corrections
\cite{Witten:1978mh}. (This is sometimes overstated
as implying that there are no quantum corrections to the
classical mass spectrum at all, while it rather means
that such quantum corrections have to affect mass and central
charge by equal amounts.) 

The simplest example of a \SUSY\ theory with
solitons, the 1+1 dimensional
\SUSY\ kink, seemed to be an exception in that the
counterparts of short and long multiplets have an equal number
of states (namely two) \cite{Witten:1978mh}
according to standard representation theory of \SUSY. Nevertheless, most of
the explicit calculations found neither nontrivial corrections
to the mass \cite{Kaul:1983yt
}
nor to the central charge \cite{Imbimbo:1984nq}.

A couple of years ago, 
this issue was reopened when two of the present authors
found \cite{Rebhan:1997iv} that the simple energy-momentum cutoff
used explicitly or implicitly in most of the calculations that obtained
a null result was inconsistent with the
integrability of
the bosonic sine-Gordon model \cite{Dashen:1975hd}.
Careful calculations using mode number cutoff such that
boundary energy contributions are avoided subsequently
established 
a nonzero result for the quantum corrections to the mass 
\cite{Nastase:1998sy,Litvintsev:2000is,Goldhaber:2000ab,Goldhaber:2002mx}
that agreed in fact with an older result by Schonfeld \cite{Schonfeld:1979hg}
and which has also been reproduced by different methods
\cite{Casahorran:1989vd,
Graham:1998qq,Bordag:2002dg}.
On the other hand, it was established by Shifman et al.\ 
\cite{Shifman:1998zy}, using a susy-preserving
higher-derivative regularization method, that there is also an
anomalous contribution to the
central charge which still leads to
BPS saturation at the quantum level\footnote{A novel
derivation of the central charge
anomaly using superspace methods to evaluate the Jacobian
in the path integral with heat-kernel methods
can be found in \cite{Fujikawa:2003gi}.}, and which was subsequently explained
by the possibility of ``super-short'' single-state supermultiplets
in 1+1 dimensions
\cite{Losev:2001uc,
Goldhaber:2000ab} which have no
definite fermion number.

In Ref.~\cite{Rebhan:2002yw} we have recently shown that
these results are most elegantly derived by employing dimensional
regularization through \SUSY-preserving dimensional reduction from
a higher-dimensional model. 
The correct quantum corrections to the mass of the \SUSY\ kink
are obtained \cite{Rebhan:2002uk}
without having to deal with energy located at boundaries
introduced in other methods, 
and the anomalous contribution to the
central charge can be obtained from corrections to the momentum
operator in the extra dimensions, which in the case of
a kink background leave a finite
remainder in the limit of 2 dimensions \cite{Rebhan:2002yw}.

In this paper we consider the
Abrikosov-Nielsen-Oleson 
\cite{Abrikosov:1957sx,Nielsen:1973cs,deVega:1976mi,Taubes:1980tm}
vortex solution of the abelian Higgs model in 2+1 dimensions
which has a supersymmetric
extension \cite{Schmidt:1992cu,Edelstein:1994bb} 
{\rm (see also \cite{Gorsky:1999hk,Vainshtein:2000hu})}
such that
classically the Bogomolnyi bound \cite{Bogomolny:1976de} is saturated.
We employ our variant
of dimensional regularization to the $N=2$ vortex by
dimensionally  reducing
the  $N=1$ abelian Higgs model in $3+1$ dimensions.
We confirm the results of 
\cite{Schmidt:1992cu,Lee:1995pm,Vassilevich:2003xk}
that in a particular gauge (background-covariant Feynman-'t Hooft)
the sums over zero-point energies of fluctuations in
the vortex background cancel completely, but 
contrary to \cite{Schmidt:1992cu,Lee:1995pm} we find
a nonvanishing quantum correction to the vortex mass
coming from a finite renormalization of the expectation value
of the Higgs field in this gauge \cite{Wimmer:2003,Vassilevich:2003xk}.
In contrast to \cite{Schmidt:1992cu}, where a null result for
the quantum corrections to the central charge was stated,
we show that the central charge receives also a net nonvanishing
quantum correction, namely from a nontrivial phase in the fluctuations
of the Higgs field in the vortex background, which
contributes to the central charge even though the latter is
a surface term that can be evaluated far away from the vortex.
The correction to the central charge exactly matches the
correction to the mass of the vortex. 

In Ref.~\cite{Lee:1995pm}, it was claimed
that the usual
multiplet shortening arguments 
in favor of BPS saturation
would not be applicable to the $N=2$ vortex since
in the vortex background there would be two
rather than one fermionic
zero modes \cite{Lee:1992yc}, leading to two short multiplets which
have the same number of states as one long multiplet.\footnote{Incidentally,
Refs.\
\cite{Lee:1995pm,Lee:1992yc} considered
supersymmetric Maxwell-Chern-Simons theory, which contains
the supersymmetric abelian Higgs model as a special case.}
We show however that the extra zero mode postulated in \cite{Lee:1995pm}
has to be discarded because its gaugino component is singular,
and that only after doing so there is agreement with the
results from index theorems \cite{Weinberg:1981eu,Lee:1992yc,Hori:2000kt}.
For this reason, standard multiplet shortening arguments do
apply, explaining the BPS saturation at the quantum level that we observe
in our explicit one-loop calculations.

\section{The vortex in $3\le D\le 4$ dimensions}

The $N=2$ \SUSY\ vortex in 2+1 dimensions is the solitonic (finite-energy) solution
of the abelian Higgs model which can be obtained
by dimensional reduction from a 3+1-dimensional $N=1$ model.
We shall use the latter for the purpose of 
dimensional regularization of the 2+1-dimensional model
by \SUSY-preserving dimensional reduction from 3+1 dimensions
(where the vortex has infinite mass but finite energy-density).

\subsection{The model}

The superspace action for the vortex in terms of 3+1-dimensional
superfields
contains
an $N=1$ abelian vector multiplet and an $N=1$ scalar multiplet,
coupled as usual, together with a Fayet-Iliopoulos term but without
superpotential,
\be
\mathcal L = 
\int d^2 \theta\, W^\alpha W_\alpha
+ \int d^4\theta\, \bar\Phi\, e^{eV} \Phi + \kappa \int d^4\theta\, V.
\ee
In terms of 2-component spinors in 3+1 
dimensions, the action reads\footnote{Our conventions
are $\eta^{\mu\nu}=(-1,+1,+1,+1)$, $\chi^{\alpha}=
\epsilon^{\alpha\beta}\chi_\beta$ and $\5\chi^{\dot\alpha}=
\epsilon^{\dot\alpha\dot\beta}\5\chi_{\dot\beta}$ with
$\epsilon^{\alpha\beta}=\epsilon_{\alpha\beta}=
-\epsilon^{\dot\alpha\dot\beta}=-\epsilon_{\dot\alpha\dot\beta}$
and $\epsilon^{12}=+1$. In particular we have
$\5\psi_{\.\alpha}=
(\psi_\alpha)^*$ but $\5\psi^{\.\alpha}=-
(\psi^\alpha)^*$. Furthermore, $\5\sigma^\mu=(-\mathbf 1,\vec\sigma)$
with the usual representation for the Pauli matrices $\vec\sigma$.}
\bea\label{L4dmodel}
\mathcal L &=& -{1\04}F_{\mu\nu}^2
+\5\chi^{\.\alpha} i \5\sigma_{\.\alpha\beta}^\mu
\6_\mu \chi^\beta +{1\02}D^2+
(\kappa-e|\phi|^2)D\nn
&&-|D_\mu \phi|^2+ \5\psi^{\.\alpha} i \5\sigma_{\.\alpha\beta}^\mu
D_\mu \psi^\beta +|F|^2 +\sqrt2 e \left[ \phi^* \chi_\alpha \psi^\alpha
+\phi \5\chi_{\.\alpha} \5\psi^{\.\alpha} \right],
\eea
where $D_\mu=\6_\mu - ieA_\mu$ when acting on $\phi$ and $\psi$,
and $F_{\mu\nu}=\6_\mu A_\nu-\6_\nu A_\mu$.
Elimination of the auxiliary field $D$ yields the scalar potential
$\mathcal V=\2 D^2=\2 e^2(|\phi|^2-v^2)^2$ with $v^2\equiv {\kappa/e}$.

In 3+1 dimensions, this model has a chiral anomaly, 
and in order to cancel the chiral U(1) anomaly, additional scalar multiplets
would be needed such that the sum over charges vanishes, $\sum_i e_i=0$.
In the present paper we shall consider only the dimensional
reduction of the minimal model (\ref{L4dmodel}), postponing
a discussion of anomaly-free 3+1 dimensional vortices 
\cite{Davis:1997bs,Gorsky:1999hk} to
a forthcoming work \cite{RVWinprep}.

In 2+1 dimensions, dimensional reduction gives an $N=2$ model involving,
in the notation of \cite{Lee:1995pm},
a real scalar $N=A_3$ and two complex (Dirac) spinors 
$\psi=(\psi^\alpha)$, $\chi=(\chi^\alpha)$.

Completing squares in the bosonic part of the classical
Hamiltonian density one finds the Bogomolnyi equations
and the central charge 
\bea
\mathcal H&=&{1\04}F_{kl}^2+|D_k \phi|^2+\2 e^2 (|\phi|^2-v^2)^2 \nn
&=& \2 |D_k \phi + i\epsilon_{kl} D_l \phi|^2
+\2 \left(F_{12}+e(|\phi|^2-v^2)\right)^2 \nn&&
+{e\02} v^2 \epsilon_{kl}F_{kl} - i \6_k (\epsilon_{kl}\phi^* D_l \phi)
\eea
where $k,l$ are the spatial indices in 2+1 dimensions.
The classical central charge reads
\be
Z=\int d^2x \, \epsilon_{kl} \6_k
\left( e v^2 A_l - i \phi^* D_l \phi \right),
\ee
where asymptotically $D_l\phi$ tends to zero exponentially fast.
Classically, BPS saturation $E=|Z|=2\pi v^2 n$ holds when
the BPS equations $(D_1 \pm iD_2)\phi \equiv D_\pm \phi=0$
and $F_{12} \pm e(|\phi|^2-v^2)=0$ are satisfied, where the
upper and lower sign corresponds to vortex and antivortex, respectively.
The vortex solution with winding number $n$ is given by
\be\label{vortexsolution}
\pv = e^{in\theta} f(r), \quad
eA_+^{\mathrm V} = -i e^{i\theta}{a(r)-n\0r},
\quad A_\pm^{\mathrm V} \equiv A_1^{\mathrm V} \pm i A_2^{\mathrm V}
\ee
where $f'(r)={a\0r}f(r)$ and $a'(r)=r e^2(f(r)^2-v^2)$
with boundary conditions \cite{Taubes:1980tm}
\bea
&a(r\to\infty)=0, &f(r\to\infty)=v,\nn
&a(r\to0)=n+O(r^2), &f(r\to0)\,\propto\, r^n+O(r^{n+2}).
\eea

\subsection{Fluctuation equations}

For the calculation of quantum corrections to a vortex solution we
decompose $\phi$ into a classical background part $\pv$ and a quantum
part $\eta$. Similarly, $A_\mu$ 
is decomposed as $A_\mu^{\mathrm V}+a_\mu$, where
only $A_\mu^{\mathrm V}$ with $\mu=1,2$ is nonvanishing.
We use a background $R_\xi$ \cite{'tHooft:1971rn
} 
gauge fixing term which is quadratic in the
quantum gauge fields,
\be\label{Lgfix}
\mathcal L_{\rm g.fix}=
-{1\02\xi} (\6_\mu a^\mu - ie\xi(\pv \eta^* - \pv^* \eta ))^2.
\ee
The corresponding Faddeev-Popov Lagrangian reads
\be
\mathcal L_{\rm ghost}=
b \left( \6_\mu^2 - e^2 \xi \left\{ 2\,|\pv|^2 + 
\pv \eta^* + \pv^* \eta  \right\} \right) c\,.
\ee

The fluctuation equations in 2+1 dimensions have been
given in \cite{Schmidt:1992cu,Lee:1995pm} for the
choice $\xi=1$ (Feynman-`t Hooft gauge
) which leads 
to important simplifications. We shall mostly use this
gauge choice when considering fluctuations in the solitonic
background, but will carry out renormalization in the
trivial vacuum for general $\xi$ to exhibit some of
the intermediate gauge dependences.

Because we are going to consider dimensional regularization
by dimensional reduction from the 3+1 dimensional model, we shall
need the form of the fluctuation equations with
derivatives in the $x^3$ direction included.
(This one trivial extra dimension will eventually be turned into
$\epsilon\to0$ dimensions.)

In the `t Hooft-Feynman gauge,
the part of the bosonic action quadratic in the quantum fields
reads 
\bea
\mathcal L_{\rm bos}^{(2)} 
&=& -\2 (\6_\mu a_\nu)^2 
- e^2 |\pv|^2 a_\mu^2 
- |D_\mu^{\mathrm V} \eta|^2 - e^2 (3|\pv|^2-v^2) |\eta|^2\nn&&
-2ie a^\mu \left[\eta^* D_\mu^{\mathrm V} \pv - \eta (D_\mu^{\mathrm V} \pv)^*
\right].\quad
\eea
In the trivial vacuum, which corresponds to $\pv\to v$ and $A_\mu^{\mathrm V}
\to 0$, the last term vanishes, but in the solitonic vacuum it
couples
the linearized field equations for the fluctuations $B\equiv(\eta,a_+/\sqrt2)$
with $a_+=a_1+ia_2$ 
to each other according to ($k=1,2$)
\be\label{Bfluceq}
(\6_3^2-\6_t^2)B=
\left(
\begin{array}{cc}
-(D_k^{\mathrm V})^2+e^2(3|\pv|^2-v^2)  &  i\sqrt2 e (D_- \pv) \\
-i\sqrt2 e (D_- \pv)^* & -\6_k^2+2 e^2 |\pv|^2
\end{array}
\right)B.
\ee
The quartet $(a_3,a_0,b,c)$ with $b,c$ the Faddeev-Popov
ghost fields has diagonal field equations at the linearized level
\be
(\6_\mu^2 - 2 e^2 |\pv|^2) Q=0, \quad Q\equiv(a_3,a_0,b,c).
\ee

For the fermionic fluctuations, which we group as
$U=\left( \psi^1 \atop \5\chi^{\.1} \right)$,
$V=\left( \psi^2 \atop \5\chi^{\.2} \right)$,
the linearized field equations
read
\be\label{UVeqs}
LU=i(\6_t+\6_3) V, \quad L^\dagger V = i(\6_t-\6_3) U,
\ee
with
\be
L=
\left(\begin{array}{cc}
iD_+^{\mathrm V} & \sqrt2 e \pv \\
-\sqrt2 e \pv^*  & i\6_-
\end{array}\right), \quad
L^\dagger =
\left(\begin{array}{cc}
iD_-^{\mathrm V} & -\sqrt2 e \pv \\
\sqrt2 e \pv^*  & i\6_+
\end{array}\right).
\ee

Iteration shows that $U$ satisfies the same second order equations
as the bosonic fluctuations $B$,
\bea
&&L^\dagger L U=(\6_3^2-\6_t^2) U, \quad L^\dagger L B=(\6_3^2-\6_t^2) B\\
&&L L^\dagger V=(\6_3^2-\6_t^2) V,
\eea
with $L^\dagger L$ given by (\ref{Bfluceq}),
whereas $V$ is governed by a diagonal equation with
\be\label{LLd}
LL^\dagger=\left(
\begin{array}{cc}
-(D_k^{\mathrm V})^2+e^2|\pv|^2+e^2 v^2  &  0 \\
0 & -\6_k^2+2 e^2 |\pv|^2
\end{array}
\right).
\ee
(In deriving these fluctuation equations we used the BPS equations throughout.)

\subsection{Renormalization}

At the classical level, the energy and central charge of vortices are
multiples of $2\pi v^2$ with $v^2=\kappa/e$. Renormalization
of tadpoles, even when only by finite amounts,
will therefore contribute directly to the 
quantum mass and central charge of the $N=2$ vortex, 
a fact that has been overlooked in the original
literature \cite{Schmidt:1992cu,Lee:1995pm}
on quantum corrections to the $N=2$ vortex.\footnote{The
nontrivial renormalization of $\kappa/e$ has however been included
in \cite{Wimmer:2003,Vassilevich:2003xk}.}

Adopting a ``minimal'' renormalization scheme where the scalar
wave function renormalization constant $Z_\phi=1$,
the renormalization of $v^2$ is fixed by the requirement
of vanishing tadpoles in the trivial sector of the 2+1 dimensional
model.
The calculation can be conveniently performed by using dimensional
regularization of the 3+1 dimensional $N=1$ model. 
For the calculation of
the tadpoles we decompose $\phi=v+\eta\equiv v+(\sigma+i\rho)/\sqrt2$, where
$\sigma$ is the Higgs field and $\rho$ the would-be Goldstone boson.
The gauge fixing term (\ref{Lgfix}) 
avoids mixed $a_\mu$-$\rho$ propagators,
but there are mixed $\chi$-$\psi$ propagators, which can be
diagonalized 
by introducing new spinors $s=(\psi+i\chi)/\sqrt2$
and $d=(\psi-i\chi)/\sqrt2$ with mass terms
$m(s_\alpha s^\alpha - d_\alpha d^\alpha)+h.c.$,
where $m=\sqrt2 e v$.

The part of the interaction Lagrangian which is relevant
for $\sigma$ tadpoles to one-loop order is given by
\be
\mathcal L^{\rm int}_{\sigma-{\rm tadpoles}}=
e(\chi_\alpha \psi^\alpha+\5\chi_{\.\alpha} \5\psi^{\.\alpha})\,\sigma
-{em\02}(\sigma^2+\rho^2)\,\sigma-em(a_\mu^2+
\xi b\,c- \delta v^2)\,\sigma, 
\ee
where $b$ and $c$ are the Faddeev-Popov fields.   

The one-loop contributions to the $\sigma$ tadpole thus read
\bea\label{sigmatadpoles}
&&\!\!\!\includegraphics[bb=0 400 350 475
]{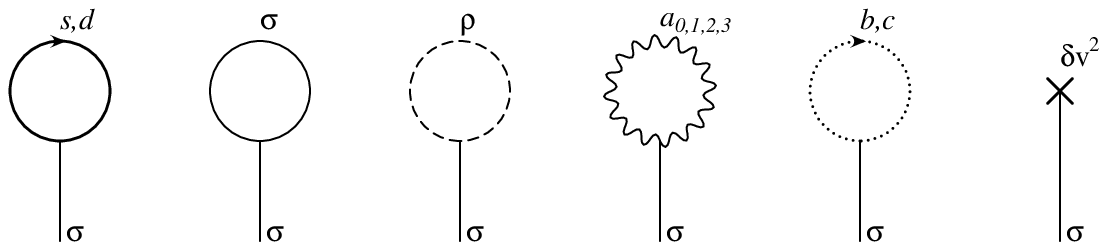}\nn
&& =(-em)\times\\
&&\!\!\!\!\!\!\!\!\!\{-2 {\rm tr}{\mathbf 1_2}I(m) + {3\02}I(m) + {1\02}I(\xi^{1\02}m) 
+ [3I(m)+\xi I(\xi^{1\02}m)] - \xi I(\xi^{1\02}m) -\delta v^2 \},\nonumber
\eea
where 
\be
I(m)=\int {d^{3+\epsilon}k\0(2\pi)^{3+\epsilon}}{-i\0k^2+m^2}
=-{m^{1+\epsilon}\0(4\pi)^{1+\epsilon/2}} {\Gamma(-\2-{\epsilon\02})\0
\Gamma(-\2)}=-{m\04\pi}+O(\epsilon).
\ee

Requiring that the sum of tadpole diagrams (\ref{sigmatadpoles}) vanishes 
fixes $\delta v^2$,
\be\label{deltav2Rxi}
\delta v^2 = \2\left(I(m)+I(\xi^{1\02}m)\right)\Big|_{D=3} = 
-{1+\xi^{1\02}\08\pi}m.
\ee
Because in dimensional regularization there are no poles in odd dimensions
at the one-loop level, the result for $\delta v^2$ is finite, but it
is nonvanishing. Because the classical mass of the vortex is
$M_{\mathrm V}=2\pi v^2 = \pi m^2/e^2$, the counterterm $\delta v^2$
is the only one that is of importance to the one-loop corrections
to $M_{\mathrm V}$. 
Since $\delta v^2$ is gauge-parameter dependent, the remaining
contributions to mass and central charge must be gauge dependent, too,
so that the final result is gauge independent.

\section{Quantum corrections to mass and central charge}

\subsection{Mass}

At the one-loop level, the quantum mass of a solitonic state is
given by
\be
M=M_{\rm cl}+\2\sum \omega_{\rm bos} - \2\sum \omega_{\rm ferm}
+\delta M
\ee
where $M_{\rm cl}$ is the classical mass expressed in terms of
renormalized parameters, $\delta M$ represents the effects
of the counter-terms to these renormalized parameters, and
the sums are over zero-point energies in the soliton background
(the zero-point energies in the trivial vacuum, which one should
subtract in principle, cancel in a \SUSY\ theory).

In the $\xi=1$ gauge
the sum over zero-point energies is formally
\bea\label{zeromodesums}
\2\sum \omega_{\rm bos} - \2\sum \omega_{\rm ferm}
&=& \sum \omega_{\eta} + \sum \omega_{a_+} - \sum \omega_U - \sum \omega_V\nn
&=& \sum \omega_U - \sum \omega_V,
\eea
where the quartet $(a_3,a_0,b,c)$ cancels separately.
(Note that in (\ref{zeromodesums}) all frequencies appear twice because
all fields are complex.)
Using dimensional regularization
as developed in \cite{Rebhan:2002uk}, these sums can be made
well defined by replacing all eigen frequencies $\omega_k$ in 2+1
dimensions by $\omega_{k,\ell}=(\omega_k^2+\ell^2)^{1/2}$ where
$\ell$ are the extra momenta, and integrating
over $\ell$. The spectral densities are
nontrivial only with respect to the 2-dimensional momenta $k$
and the former are not modified by dimensional regularization.

In \cite{Schmidt:1992cu,Lee:1995pm} it has been shown that
$L^\dagger L$ and $LL^\dagger$,
which govern $U$ and $V$, respectively,
are isospectral up to zero modes.
In dimensional regularization, where the zero-mode
contributions continue to give zero because scaleless
integrals vanish, one can therefore conclude that in the $\xi=1$ gauge
there is a complete cancellation of the sums over zero-point
energies. All that remains is the finite renormalization of $\delta v^2$
in that gauge:
\be\label{MV}
E
= 2\pi |n| (v^2 + \delta v^2|_{\xi=1}) = 
2\pi |n| (v^2 - {m\04\pi})
\equiv |n| \left( { \pi m^2 \0 e^2} - {m\02} \right)
\ee
(In gauges other than $\xi=1$ the fluctuation equations for the
$B$ fields, i.e.\ $\eta,a_+$, no longer match those of the $U$ fermions.)

This result agrees with \cite{Vassilevich:2003xk}, where however
a careful analysis of boundary conditions in the heat-kernel
approach was needed because the vortex had to be put in a box
to discretize the spectrum. In dimensional regularization one does
not need to put the system in a box, and as a consequence there is
no need to study the contributions from these artificial
boundaries.

\subsection{Central charge}

By starting from the \SUSY\ algebra in 3+1 dimensions
one can derive the central charge in 2+1 dimensions as the
component $T_{03}$ of
\be
T^{\mu\nu}=-{i\04} {\rm Tr} \,\sigma^{\mu \alpha\dot\alpha}\,
\{ \bar Q_{\dot\alpha},J_\alpha^\nu\}
\ee
where $J_\alpha^\nu$ is the \SUSY\ Noether current.

The antisymmetric part of $T^{\mu\nu}$ gives the
standard expression for the central charge density, while
the symmetric part is a genuine momentum in the extra
dimension:
\be
\< Z \> = \int d^2 x \< T^{03} \> 
= \< \tilde Z + \tilde P_3 \>.
\ee
(A similar decomposition is valid for the kink \cite{Rebhan:2002yw}.)

$\tilde Z$ corresponds to the classical expression for the central
charge. Being a surface term, its quantum corrections can be
evaluated at infinity:
\be
\langle \tilde Z \rangle =
\int d^2x \6_k \epsilon_{kl} \langle \tilde\zeta_l \rangle
= \int_0^{2\pi}\!\!\! d\theta \langle \tilde\zeta_\theta \rangle|_{r \to \infty}
\ee
with 
$\tilde\zeta_l=e v^2_0 A_l-i\phi^\dagger D_l \phi$ and $v_0^2=v^2+\delta v^2$.

Expanding in quantum fields $\phi=\pv+\eta$, 
$A=A^{\rm V}+a$ and using that
the classical fields $\pv\to v e^{in\theta}$, $A_\theta^{\rm V}\to n/e$,
$D_\theta^{\rm V} \pv \to 0$ as $r\to \infty$,
we 
obtain to one-loop order 
\bea\label{Zab}
&&\langle \tilde Z \rangle 
= 2\pi n v^2_0 - i \int_0^{2\pi}\!\!\! d\theta \left\langle 
(\pv^*+\eta^\dagger) (D_\theta^{\rm V}-ie a_\theta)(\pv+\eta)
\right\rangle|_{r \to \infty} \nn
&=& 2\pi n \{v^2_0 - \langle \eta^\dagger \eta \rangle |_{r \to \infty}\}
- i \int_0^{2\pi}\!\!\! d\theta \left\{
\left\langle 
\eta^\dagger \6_\theta \eta 
\right\rangle
-ie \pv^* \left\langle a_\theta \eta \right\rangle
-ie \pv \left\langle a_\theta \eta^\dagger \right\rangle
\right\}|_{r \to \infty}\nn
&\equiv& Z_a+Z_b
\eea
where we have used 
$\langle \eta(r\!\to\!\infty) \rangle \to 0$ (which determines $\delta v^2$),
$\int_0^{2\pi}\! d\theta\langle a_\theta \rangle = 0$, and
$\langle \eta^\dagger \eta a_\theta \rangle = O(\hbar^2)$.

The first contribution, $Z_a$, can be easily evaluated for
arbitrary gauge parameter $\xi$, yielding
\bea
Z_a&=&2\pi n \{v^2_0
-\2(
\langle \sigma\sigma \rangle + \langle \rho\rho \rangle) |_{r \to \infty}\}
\nn&=&
2\pi n \{v^2_0-\2[I(m)+I(\xi^{1\02}m)]\}\nn&=&2\pi n (v_0^2-\delta v^2)=
2\pi n v^2.
\eea
If this was all, the quantum corrections to $Z$
would cancel, just as in the naive
calculation of $Z$ in the \SUSY\ kink \cite{Imbimbo:1984nq,Rebhan:1997iv}.

The second contribution in (\ref{Zab}), however, does not
vanish when taking the limit $r\to\infty$. This contribution
is simplest in the $\xi=1$ gauge, where the $\eta$ and $a_\theta$
fluctuations are governed by the fluctuation equations
(\ref{Bfluceq}). In the limit $r\to\infty$ one has
$|\pv|\to v$ and $D_-\pv\to 0$ exponentially. This
eliminates the contributions from $\left\langle a_\theta \eta \right\rangle$.
However, $D_k^2$, which governs the $\eta$ fluctuations,
contains long-range contributions from the vector
potential. Making a separation of variables in $r$ and $\theta$
one finds that asymptotically 
\be
|D_k^{\mathrm V} \eta|^2 \to |\6_r \eta|^2+{1\0r^2} |(\6_\theta-in)\eta|^2
\ee
so that the $\eta$ fluctuations
have an extra phase factor $e^{in\theta}$ compared to those in the
trivial vacuum. We thus have, in the $\xi=1$ gauge,
\bea
Z_b&=&
- i \int_0^{2\pi}\!\!\! d\theta \left\{
\left\langle 
\eta^\dagger \6_\theta \eta 
\right\rangle
-ie \pv^* \left\langle a_\theta \eta \right\rangle
-ie \pv \left\langle a_\theta \eta^\dagger \right\rangle
\right\}|_{r \to \infty}\nn
&=&- i \int_0^{2\pi}\!\!\! d\theta 
\left\langle \eta^\dagger \6_\theta \eta 
\right\rangle_{\xi=1} 
= 2\pi n \left\langle \eta^\dagger \eta \right\rangle_{\xi=1,r \to \infty}
= 2\pi n\, \delta v^2\Big|_{\xi=1}\,.
\eea
This is exactly the result for the one-loop correction to the
mass of the vortex in
eq.~(\ref{MV}), implying saturation of the BPS bound 
provided that there are now no anomalous contributions to the
central charge operator 
as there are in the case in the $N=1$ \SUSY\
kink \cite{Rebhan:2002yw}.

In 
dimensional regularization by dimensional
reduction from a higher-dimensional model such anomalous
contributions to the central charge operator come from
a finite remainder of the extra momentum operator \cite{Rebhan:2002yw}. 
The latter is
given by \cite{Lee:1995pm}
\bea
Z_c=\< \tilde P_3\>&\!=\!&
\int d^{2}x \< F_{0i}F_{3i}
+ (D_0 \phi)^\dagger D_3 \phi
+ (D_3 \phi)^\dagger D_0 \phi \right. \nn
&&\qquad\qquad\left.
- i \bar \chi \bar\sigma_0 \6_3 \chi
-i \bar \psi \bar\sigma_0 D_3 \psi \>.
\eea
Inserting mode expansions for the quantum fields one
immediately finds that the bosonic contributions vanish 
because of symmetry in the extra trivial dimension.
However, this is not the case for the fermionic fields,
which have a mode expansion of the form
\be
\left( U \atop V \right) =
\int{d^\epsilon \ell \0 (2\pi)^{\epsilon/2}}
{\Sigma}\!\!\!\!\!\!\int_k
{1\0\sqrt{2\omega}} \Bigl\{
b_{k,\ell}\, e^{-i(\omega t-\ell z)}
\left(  {\sqrt{\omega-\ell}\, u_1 \atop \sqrt{\omega-\ell}\, u_2} 
\atop   {\sqrt{\omega+\ell}\,v_1 \atop \sqrt{\omega+\ell}\,v_2} \right)
+d_{k,\ell}^\dagger \times(c.c.)
\Bigr\},
\ee
where we have not written out explicitly the zero-modes
(for which $\omega^2=\ell^2$).
The fermionic contribution to $Z_c$ reads, schematically,
\bea
Z_c&=&\< \tilde P_3\>=\nn
&=& \int{d^\epsilon \ell \0 (2\pi)^\epsilon}
{\Sigma}\!\!\!\!\!\!\int_k {\ell^2\02\omega}
\int d^2x \left[ |u_1|^2+|u_2|^2-|v_1|^2-|v_2|^2 \right]
\eea
where $\omega=\sqrt{\omega_k+\ell^2}$, so that the $\ell$ integral
does give a nonvanishing result. However, the $x$-integration
over the mode functions $u_{1,2}(k;x)$ and $v_{1,2}(k;x)$
produces their spectral densities, which cancel up to
zero-mode contributions as we have
seen above\footnote{An explicit calculation which
confirms these cancellations will be presented in
\cite{RVWinprep}.}, and zero-mode contributions
only produce scaleless integrals
which vanish in dimensional regularization. Hence, $Z_c=0$
and $|Z|=|Z_a+Z_b|=E$, so that the BPS bound is saturated at
the (one-loop) quantum level.

\subsection{Fermionic zero modes and multiplet shortening}

Massive representations of the Poincar\'e supersymmetry algebra 
for which the absolute value
of the central charge equals the energy, i.e.~when the BPS bound is
saturated, contain as many states as 
massless representations, which is half of that of
massive representations for which the BPS bound is not saturated. 
These results
also apply in 2+1 dimensions for the $N=2$ super-Poincar\'e algebra
\cite{Lee:1995pm}.

A particular multiplet of states is obtained by taking the vortex solution,
and acting on it with the \SUSY\ generators of the $N=2$ \SUSY\ algebra,
which contains two complex charges $Q_+$, $Q_-$, and their hermitian
conjugates $(Q_+)^\dagger$ and $(Q_-)^\dagger$. One of these charges,
$Q_-$, annihilates the vortex solution, while the other one, $Q_+$, is to
linear order in quantum operators proportional to the annihilation
operator of a fermionic zero mode. 

However, if there indeed is a second fermionic
zero mode in the model as postulated in \cite{Lee:1995pm}\footnote{\rm
In the literature one can in fact find two different conventions for
indicating the number of fermionic zero modes.
Like Refs.~\cite{Lee:1995pm,Hori:2000kt} we only count the number of zero
modes in the fermionic quartet $(U,V)$ and not additionally
those in the corresponding conjugated fields $(U^\dagger,V^\dagger)$.
One zero mode in this way of counting then corresponds to
a {\em pair} of creation/annihilation operators. Alternatively
one may count the zero modes in both $(U,V)$ and $(U^\dagger,V^\dagger)$
and thus ascribe one zero mode
to each creation or annihilation operator.
The latter way of counting is
employed for instance in Ref.~\cite{Vainshtein:2000hu}.}, 
in second quantization it would be present in the
mode expansion of the fermionic quartet $U$ and $V$,
\be
\left(U \atop V\right)
=a_{\mathrm I} 
\left( {\psi_{\mathrm I}^1 \atop \bar\chi^{\dot 1}_{\mathrm I}} \atop {0 \atop 0}\right)
+a_{\mathrm{I\!I}} 
\left( {\psi_{\mathrm{I\!I}}^1 \atop \bar\chi^{\dot 1}_{\mathrm{I\!I}}} \atop {0 \atop 0}\right)+
\mbox{non-zero modes.}
\ee
As a result, there would then be a quartet of BPS states
\be\label{quartet}
\ket{v},\;
a_{\mathrm I}^\dagger\ket{v},\;
a_{\mathrm{I\!I}}^\dagger\ket{v},\;
a_{\mathrm I}^\dagger a_{\mathrm{I\!I}}^\dagger\ket{v}
\ee
comprising two short multiplets of $N=2$ \SUSY, which
are degenerate and together have as many states as one
long multiplet without BPS saturation.  As stressed in \cite{Lee:1995pm},
the standard argument
for stability of BPS saturation under quantum corrections
from multiplet shortening \cite{Witten:1978mh} thus would not be applicable.

However, we shall now show that there is in fact only a
single fermionic zero mode in a vortex background with
winding number $n=1$.
To this end,
we first observe that the zero modes must lie
in $U$, because $V$ is governed by the operator $LL^\dagger$ of
Eq.~(\ref{LLd}), whose only zero mode solution is $V_0\equiv 0$.
A zero mode for $U$ must satisfy $LU=0$, 
and to analyse this equation we follow \cite{Lee:1995pm} and
set $\psi^1(x,y)=-i e^{i(j-\2+n)\theta}
u(r)$ and $\bar\chi^{\dot 1}=e^{i(j+\2)\theta}d(r)$.
The equation $LU=0$ reduces then to
\be\label{udeqs}
\left(\begin{array}{cc}
\6_r-{a+j-\2\0r} & \sqrt2 e f \\
\sqrt2 e f & \6_r+{j+\2\0r}
\end{array}\right)
\Biggl({u\atop d}\Biggr) = 0,
\ee
where $f=f(r)$ and $a=a(r)$ satisfy $f'={a\0r}f$ and $a'=re^2(f^2-e^2)$.
Iterating this equation yields
\be\label{uzm}
\left( \6_r^2 + {1\0r}\6_r - {(j-\2)^2\0r^2} - 2 e^2 f^2 \right)
{u\0f} = 0.
\ee
Given a solution for $u$, the corresponding solution for $d$ follows from
$LU=0$.

For given $j$, this equation has two independent solutions, a linear
combination of which yields solutions which decrease exponentially
fast as $r\to\infty$. Hence, {\em both} solutions should be regular
at $r=0$. For $j\not=\2$, one has, using $f(r\to0)\sim r^n$,
\be
\psi^1 \sim u\sim r^n(C_1 r^{j-\2} + C_2 r^{-(j-\2)}) \quad
\mbox{for $r\to 0$}
\ee
which selects for $n=1$ only $j=-\2$. This solution is the zero mode
that is  obtained by acting with
$Q_+$ on the background solutions,
which gives $\psi^1=-i D_-\phi_{\rm V}/\sqrt2=-i\sqrt2 f'$,
$\bar\chi^{\dot1}=F_{12}=-e(f^2-v^2)$. For $j=\2$, one finds for $n=1$ near
$r=0$
\be\label{seczmorig}
\psi^1 \sim C_1\,(x+iy)+C_2\,(x+iy)\ln r\;.
\ee
For large $r$, $\psi_1 \sim e^{-mr}e^{i\theta}$, as follows from
(\ref{uzm}).
This solution corresponds to the second fermionic zero mode
postulated in Ref.~\cite{Lee:1995pm}.

However, while (\ref{seczmorig}) is regular at the origin,
the associated gaugino component is not: (\ref{udeqs}) implies
that
\be
\bar\chi^{\dot 1}\sim C_2 {e^{i\theta}\0r},
\ee
so this solution has to be discarded when $C_2\not=0$.

Similarly, one can show that for winding number $n>1$ regularity
of the gaugino component generically requires that $j\le-\2$ so that
the correct quantization condition for normalizable fermionic zero modes
is $-n+\2 \le j \le -\2$. Hence, there are $n$ independent
fermionic zero modes, not $2n$ as concluded in \cite{Lee:1995pm}.
It is in fact only the former value that agrees with the
results \cite{Lee:1992yc,Hori:2000kt} 
obtained from the index theorem 
\cite{Weinberg:1981eu}.

{\rm
As has been proved rigorously in \cite{Weinberg:1979er},
in the bosonic sector there are $2n$ zero modes,
which are related to the above
$n$ independent fermionic zero modes by supersymmetry.
In the $R_{\xi=1}$ background gauge 
$\6_\mu a^\mu - ie(\pv \eta^* - \pv^* \eta )=0$,
the bosonic zero modes 
satisfy a set of equations completely equivalent
to those for the fermionic zero modes \cite{Lee:1992yc}.
But
the linearly dependent solutions $({U \atop 0})$ and $i({U \atop 0})$
correspond to linearly independent solutions for the bosonic
zero modes $a$ and $\eta$.\footnote{\rm For an analogous case 
see eq.\ (3.8) of 
Ref.~\cite{Weinberg:1979ma}.} 
In particular, for $n=1$, the
$j=-\2$ solution
$(\psi^1=-iu(r),\bar\chi^{\dot 1}=d(r))$ with real $u(r)$ and $d(r)$
corresponds to the bosonic zero mode $\eta(r)=-iu(r)$,
$(a_1,a_2)=(\sqrt2 d(r),0)$, while multiplying the fermionic
solution by $i$ corresponds to the bosonic zero mode
$\eta(r)=u(r)$, $(a_1,a_2)=(0,\sqrt2 d(r))$, which is evidently
linearly independent of the former. 
For both solutions the $R_{\xi=1}$ gauge condition is satisfied due to
the lower component of the field equation (\ref{udeqs}).
Conversely, one can start from the classical vortex solution
and find two independent bosonic zero modes by considering
their derivatives with respect to the $x$ and $y$ coordinates.
Performing a gauge transformation to satisfy the
$R_{\xi=1}$ gauge condition leads one back to the above solutions.
This additionally confirms that the above analysis has
identified all fermionic zero modes in the quartet $(U,V)$.
}

We thus conclude that for the basic vortex (winding number $n=1$)
there is exactly one fermionic zero mode ({\rm corresponding to
one pair of fermionic creation/annihilation operators})
and this gives rise
to a single short multiplet at the quantum level.
Standard multiplet shortening arguments therefore do apply and
explain the preservation of BPS saturation that we verified at
one-loop order.
 
\section*{Acknowledgments}

We would like to thank Luzi Bergamin, Cumrun Vafa, 
Arkady Vainshtein, and Dima Vassilevich for
discussions.
P.v.N. thanks the Erwin-Schr\"odinger-Institute
for financing a stay of his in Vienna. R.W.\ has been
supported by the Austrian Science Foundation FWF, project
no. P15449.

\small


\end{document}